\documentclass{elsart}
\usepackage{natbib,amsmath,amssymb,graphicx,latexsym} 
\def\beq{\nopagebreak \begin{equation}}
\def\eeq{\end{equation}}
\def\bk{{\bf k}}
\def\bR{{\bf R}}
\def\df{\Bigl[-\frac{\partial f_\mu(T;\varepsilon)}{\partial \varepsilon} \Bigr]}
\def\bmpt#1{\begin{minipage}[t]{#1\linewidth} \vspace{0pt}}
\def\emp{\end{minipage}}

\newcounter{bla}

\begin{document}
\begin{frontmatter}

\title{BoltzTraP. A code for calculating band-structure dependent quantities.}
\author[auchem]{Georg K. H. Madsen\corauthref{corauthor}},
%\author[auphys]{Gilles Santi}
\corauth[corauthor]{Corresponding author}
\ead{georg@chem.au.dk}
\author[nrl]{David J. Singh},

\address[auchem]{Dept. of Chemistry, University of Aarhus, DK-8000 \AA rhus C, Denmark.}
\address[nrl]{Condensed Matter Sciences Division, Oak Ridge National Laboratory, Oak Ridge, TN 37831-6032, U.S.A.}

\begin{abstract}
A program for calculating the semi-classic transport coefficients is described. It is based on a smoothed Fourier interpolation of the bands. From this analytical representation we calculate the derivatives necessary for the transport distributions. The method is compared to earlier calculations, which in principle should be exact within Boltzmann theory, and a very convincing agreement is found.

\begin{flushleft}
  %Insert your suggested PACS number here

\end{flushleft}


\end{abstract}

\end{frontmatter}

% Computer program descriptions should contain the following
% PROGRAM SUMMARY.

{\bf PROGRAM SUMMARY}
  %Delete as appropriate.

\begin{small}
\noindent
{\em Manuscript Title:} BoltzTraP. A code for calculating band-structure dependent quantities.                                        \\
{\em Authors:} Georg K. H. Madsen, David J. Singh                                             \\
{\em Program Title:}   BoltzTrap                                       \\
{\em Journal Reference:}                                      \\
  %Leave blank, supplied by Elsevier.
{\em Catalogue identifier:}                                   \\
  %Leave blank, supplied by Elsevier.
{\em Licensing provisions:}     none                              \\
  %enter "none" if CPC non-profit use license is sufficient.
{\em Programming language:}     Fortran90                              \\
{\em Computer:}                 The program should work on any system with a F90 compiler. The code has been tested with the Intel Fortran compiler.      \\
  %Computer(s) for which program has been designed.
{\em Operating system:}         Unix/Linux                              \\
  %Operating system(s) for which program has been designed.
{\em RAM:} bytes                   Up to 2 Gb for low symmetry, small unit cell structures                           \\
  %RAM in bytes required to execute program with typical data.
{\em Classification:}     7.9 Transport Properties                            \\
  %Classify using CPC Program Library Subject Index
  %e.g. 4.4 Feynman diagrams, 5 Computer Algebra.
{\em External routines:}  The LaPack and Blas libraries are needed                                    \\
  % Fill in if necessary, otherwise leave out.
{\em Nature of problem:}Analytic expansion of energy-bands. Calculation of semi-classic integrals \\
  %Describe the nature of the problem here.
{\em Solution method:}Smoothed Fourier expansion of bands.\\
{\em Running time:}\\
Up to 3 hours for low symmetry, small unit cell structures\\
  %Give an indication of the typical running time here.
   \\
\end{small}

\newpage

% In program descriptions the main text of the paper is listed under
% the heading LONG WRITE-UP.

\hspace{1pc}
{\bf LONG WRITE-UP}

\section{Introduction}
Method developments, the existence of user friendly distributed codes and the ever increasing computer power are making the calculation of band-structures, for even relatively complex materials, more and more straight forward. As several properties can be calculated from the energy bands and their derivatives, the usefulness off a generally applicable, easily portable and documented code for analysis of the bands should be clear.

The code presented here relies on a Fourier expansion of the band energies where the space group symmetry is maintained by using star functions. The idea of the Fourier expansion is to use more star functions than band energies, but to constrain the fit so that the extrapolated energies are exactly equal to the calculated band-energies and use the additional freedom to minimize a roughness function  and thereby suppress oscillations between the data-points.\cite{smooba2,smooba1,pickett88} Using the analytical representation of the bands it is then a reasonable simple procedure to calculate band-structure dependent quantities. The method has been tested for several applications based on Boltzmann theory, including the transport coefficients of intermetallic compounds,\cite{allenim} high $T_C$ superconductors\cite{allen88} and thermoelectrics.\cite{davidskut2} Furthermore the present code has already been applied to calculate the transport coefficients in a series of different clathrate structures\cite{gmeugage} and a very good agreement was found with experimental values.\cite{bentien05_1,bentien05_2} The good agreement was also found for the demanding Hall coefficient that depends on the second derivative of the bands.\cite{gmeugage,bentien05_2}

Because of the known limitations of Boltzmann theory\cite{allenrev} the comparison with experimental measurements is not the best method for testing the actual algorithm for expanding the bands. As the interpolated bands pass exactly though the calculated band energies, the precision of the present method is mainly limited by possible band crossings where the band derivatives will be calculated wrongly. We will therefore test our method by comparing with the resent results by Scheidemantel et al.\cite{sofote}. Scheidemantel et al. calculated transport coefficients of Bi$_2$Te$_3$\cite{sofote} by calculating the group velocities from the momentum matrix elements. As the momentum matrix elements can be calculated directly from the wavefunction,\cite{cadoptic} their method should avoid any problems at band crossings.\cite{sofote} As the calculations were documented in detail\cite{sofote} and Bi$_2$Te$_3$ has a complex band-structure that is strongly influenced by spin orbit coupling, it constitutes a challenging test-case which we will use in the present paper.

\section{Code implementation}
\subsection{Algorithms}
The code relies on a Fourier expansion of the band energies where the space group symmetry is maintained by using star functions
\beq
\tilde{\varepsilon}_i(\bk)=\sum_{\bf R} c_{\bR i} S_\bR(\bk) \quad , \quad S_\bR(\bk)=\frac{1}{n}\sum_{\{\Lambda\}} e^{i \bk \cdot \Lambda\bR}
\label{eq:fourexp}
\eeq
where \bR\ is a direct lattice vector,  ${\{\Lambda\}}$ are the $n$ point group rotations. The idea of the Fourier expansion is to use more star functions than band energies, but to constrain the fit so $\tilde{\varepsilon}_i$ are exactly equal to the band-energies, $\varepsilon_i$ and use the additional freedom to minimize a roughness function.\cite{smooba2,smooba1,pickett88} The choice of the roughness function, $\rho_\bR$, was discussed by Pickett et al.\cite{pickett88} who found the following expression to be useful for suppressing oscillations between the data-points.
\beq
\rho_\bR=\left(1-C_1\left(\frac{|\bR|}{|\bR_{min}|}\right)^2\right)^2+C_2\left(\frac{|\bR|}{|\bR_{min}|}\right)^6
\eeq
where $\bR_{min}$ is smallest nonzero lattice vector. $C_1$ and $C_2$ are parameters, but our and earlier\cite{pickett88} experience found that the results are quite insensitive to their actual value and we have therefore fixed them to $C_1=C_2=3/4$. To ensure that $\tilde{\varepsilon}_i$ pass exactly through the calculated the band energies at the same time as the roughness function is minimized, the algorithm needs sufficient freedom. This means that the number of planewaves must be larger than the number of band energies. The number of planewaves to the number of band-energies is controlled by the input parameter \texttt{LPFAC}, Table~\ref{tab:input}, and the program prints a warning if the fit is poor (subroutine KCOMP).

The expansion coefficients are given
\beq
c_{\bR i}=\begin{cases}
\varepsilon_i(\bk_N)-\sum_{{\bf R}\neq {\bf 0}} \frac1{n_{\bf R}} c_{{\bf R} i} e^{i \bk \cdot {\bf R}} & \bR={\bf 0} \\
\rho^{-1}_\bR\sum_{\bk \neq \bk_N} \lambda_\bk [S^*_{\bR i}-S^*_{0 i}] & \bR \neq {\bf 0} \end{cases}
\eeq
where $\lambda_\bR$ are calculated by solving 
\beq
\Delta \varepsilon_i(\bk) = \varepsilon_i(\bk)-\varepsilon_i(\bk_N)=\sum_{\bk' \neq \bk_N} H_{\bk \bk'} \lambda_\bR^i 
\label{eq:coef1}
\eeq
where
\beq
 H_{\bk \bk'}=\sum_{\bR \neq {\bf 0}} \frac{[S_\bR(\bk)-S_\bR(\bk_N)][S^*_\bR(\bk')-S^*_\bR(\bk_N)]}{\rho_\bR} 
\label{eq:coef2}
\eeq
The time consuming steps in the Fourier expansion are obviously the solution of Eq.~(\ref{eq:coef1}) and the setup of the $H_{\bk \bk'}$ matrix, Eq.~(\ref{eq:coef2}). The matrix setup can be identified as a multiplication of two $k\times R$ matrices and both Eq.~(\ref{eq:coef1}) and Eq.~(\ref{eq:coef2}) can thus be handled efficiently by BLAS calls.\cite{BLAS,LAPACK} The calculation of the expansion parameters, $c_{\bR,i}$, are carried out in the subroutine FITE4.

\section{Test problems}
\subsection{Boltzmann theory: The semi-classic equations}
Boltzmann theory\cite{allenrev,zimanelpho,hurdhall} is a useful tool for gaining insight into the transport properties of real materials. In the presence of an electric and magnetic field and a thermal gradient the electric current, $j$, can be written in terms of the conductivity tensors
\begin{equation}
j_i=\sigma_{ij}E_j+\sigma_{ijk} E_j B_k + \nu_{ij}\nabla\!_j T +\cdots
\label{eq:etrans}
\end{equation}

In terms of the group velocity
\begin{equation}
v_\alpha(i,\bk)=\frac{1}{\hbar}\frac{\partial\varepsilon_{i,{\bf k}}}{\partial k_\alpha}
\label{eq:fermideriv}
\eeq
and the inverse mass tensor
%\footnote{
%Sometimes conductivity is discussed in terms of the effective mass tensor
%\begin{equation*}
%\Bigl(\frac{n}{m}\Bigr)_{\alpha\beta}(i,\bk)=\frac{1}{\Omega} v_\alpha(i,{\bf k}) v_\beta(i,{\bf k}) 
%\end{equation*}
%so the conductivity is given as $\sigma_{\alpha\beta}=e^2\tau_{i,{\bf k}}(n/m)_{\alpha\beta}$. We will not use this notation, to avoid confusion with the inverse mass tensor.
%}
\begin{equation}
M^{-1}_{\beta u}(i,\bk)=\frac{1}{\hbar^2}\frac{\partial^2\varepsilon_{i,{\bf k}}}{\partial k_\beta \partial k_u} 
\label{eq:invmass}
\end{equation}
the conductivity tensors can be obtained
\beq
\sigma_{\alpha\beta}(i,{\bf k})=e^2\tau_{i,{\bf k}} v_\alpha(i,{\bf k}) v_\beta(i,{\bf k}) \label{eq:sigxx} 
\eeq
while $\sigma_{\alpha\beta\gamma}$ is most elegantly written using the Levi-Civita symbol, $\epsilon_{ijk}$\cite{hurdhall,arfken}
\beq
\sigma_{\alpha\beta\gamma}(i,{\bf k})=e^3\tau^2_{i,{\bf k}} \epsilon_{\gamma u v} v_\alpha(i,{\bf k})v_v(i,{\bf k}) M^{-1}_{\beta u} 
\label{eq:sigxxx}
\eeq

The notation used in Eqs.~(\ref{eq:sigxx}-\ref{eq:sigxxx}) gives directly the symmetry of the conductivity tensors. F.inst. in an orthorhombic symmetry $\sigma_{\alpha\beta}$ is diagonal with all three components independent and $\sigma_{\alpha\beta\gamma}$ has three independent components and vanishes unless $\alpha$,$\beta$ and $\gamma$ are all different.

The relaxation time, $\tau$, in principle is dependant on both the band index and the $\bk$ vector direction. However detailed studies of the direction dependence of $\tau$ have shown that, to a good approximation, $\tau$ is direction independent\cite{schulzhall} and that even in the superconducting cuprates, that have substantially anisotropic conduction and cell-axes, the $\tau$ is almost isotropic.\cite{allen88} In the present we will use the simplest approximation for the relaxation time, namely to keep it constant, which is the most often used in praxis.

Similar to the density of states energy projected conductivity tensors can be defined using the conductivity tensors, Eqs.~(\ref{eq:sigxx}-\ref{eq:sigxxx})
\begin{equation}
\sigma_{\alpha\beta}(\varepsilon)= \frac{1}{N}\sum_{i,{\bf k}} \sigma_{\alpha\beta}(i,{\bf k}) \frac{\delta(\varepsilon-\varepsilon_{i,{\bf k}})}{d\varepsilon}
\label{eq:transdist}
\end{equation}
where $N$ is the number of ${\bk}$-points sampled. Similarly $\sigma_{\alpha\beta\gamma}(\varepsilon)$ can be defined. With the expansion of the bands, Eq.~(\ref{eq:fourexp}), the necessary derivatives, Eq.~(\ref{eq:fermideriv}), are straightforwardly calculated as Fourier sums which can be efficiently evaluated using fast Fourier transforms (FFTs). Evaluation of the density of states and transport distributions thus requires a total of 10 FFTs for each band in the general case. The calculation of the transport distributions is carried out in the subroutine DOS and are output to the files: case.transdos, case.sigxx, case.sigxxx. 

The transport tensors, Eq.~(\ref{eq:etrans}), can then be calculated from the conductivity distributions
\begin{gather}
\sigma_{\alpha\beta}(T;\mu)=\frac{1}{\Omega}\int\! \sigma_{\alpha\beta}(\varepsilon) \df d\varepsilon \label{eq:Isigxx} \\
\nu_{\alpha\beta}(T;\mu)=\frac{1}{eT\Omega} \int\! \sigma_{\alpha\beta}(\varepsilon) (\varepsilon-\mu)\df d\varepsilon \\
\kappa^0_{\alpha\beta}(T;\mu)=\frac{1}{e^2T\Omega}\int\! \sigma_{\alpha\beta}(\varepsilon) (\varepsilon-\mu)^2 \df d\varepsilon \label{eq:k0} \\
\sigma_{\alpha\beta\gamma}(T;\mu)=\frac{1}{\Omega}\int\! \sigma_{\alpha\beta\gamma}(\varepsilon) \df d\varepsilon 
\label{eq:Isigxxx}
\end{gather}
where $\kappa^0$ is the electronic part of the thermal conductivity. The Seebeck and Hall coefficients can then easily be calculated
\begin{gather}
S_{ij}=E_i(\nabla\!_j T)^{-1}=(\sigma^{-1})_{\alpha i}\nu_{\alpha j} \\
R_{ijk}=\frac{E^{ind}_j}{j^{appl}_i B^{appl}_k} = (\sigma^{-1})_{\alpha j} \sigma_{\alpha\beta k}(\sigma^{-1})_{i\beta}
\end{gather}
Under the assumption that the relaxation time $\tau$ is direction independent, both the Seebeck and the Hall coefficients are independent of $\tau$. The integrals Eqs.~(\ref{eq:Isigxx})-(\ref{eq:Isigxxx}) are performed in the subroutine FERMIINTEGRALS.

\subsubsection{Test case: Bi$_2$Te$_3$}
\begin{figure}[t]
\bmpt{.02}
(a)
\emp
\bmpt{.46}
\includegraphics[width=\linewidth]{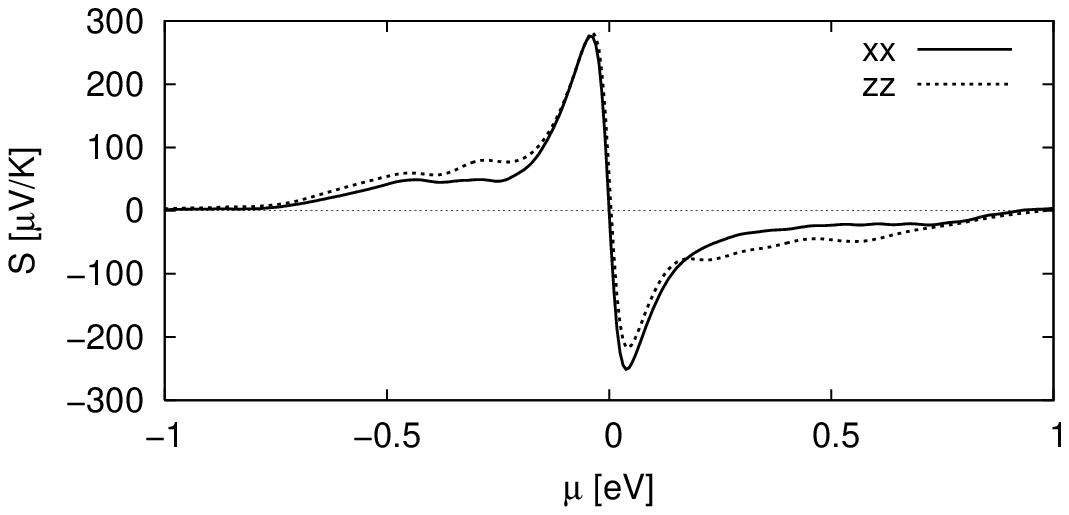} 
\emp
\bmpt{.02}
(b)
\emp
\bmpt{.46}
\includegraphics[width=\linewidth]{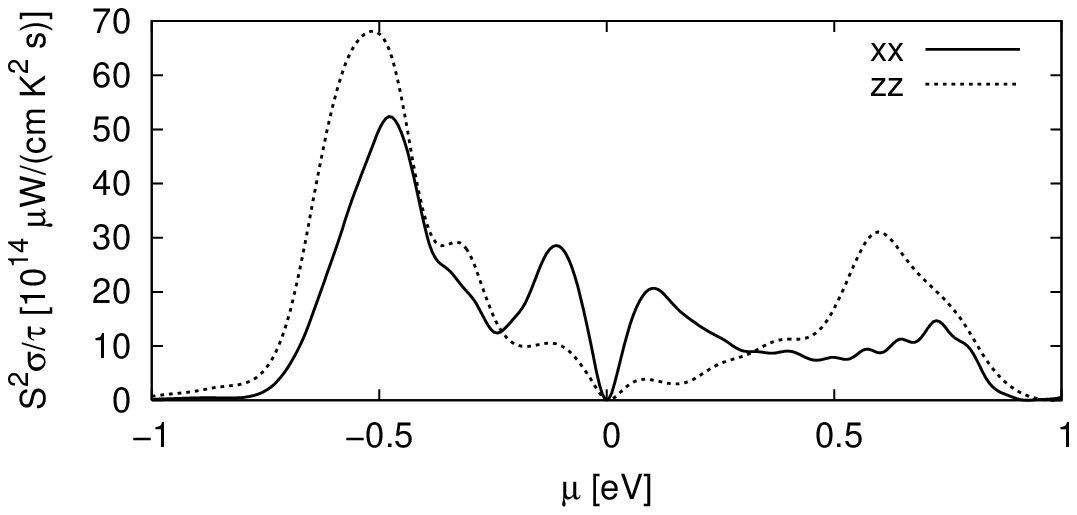} 
\emp
\caption{Transport coefficients as a function of chemical potential: (a) Seebeck coefficient (b) Power factor with respect to scattering time $S^2\sigma/\tau$. One obtains the power factor in the usual units of $\mu$W/(cm K$^2$) by multiplying by $\tau$ in units of $10^{-14}$~s.}
\label{fig:BiTe}
\end{figure}

The calculation was carried out using the WIEN code\cite{wien2k} with the same computational parameters as in ref.~\cite{sofote}. The calculated transport coefficients were found to be converged using a non-shifted mesh with 56000~k points (4960 in the IBZ). The original $k$-mesh was interpolated onto a mesh four times as dense. 

The calculated transport coefficients are given in Figure~\ref{fig:BiTe}. Figure~\ref{fig:BiTe}a,b show the Seebeck coefficient and the power factor with respect to scattering time. Both these curves can be compared to the earlier work\cite{sofote} and an excellent agreement is found, both with respect to shape and absolute values. 

\begin{figure}[t]
\center
\includegraphics[width=.5\linewidth]{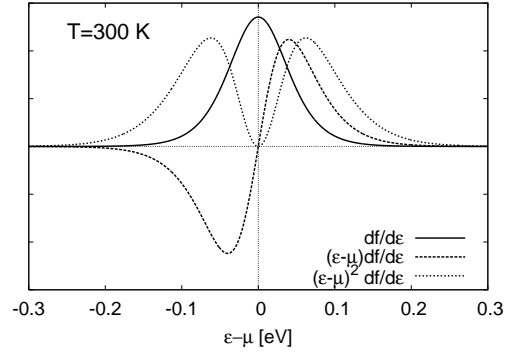} 
\caption{Integrand factor $\partial f/\partial \varepsilon$ in Eqs.~(\ref{eq:Isigxx})-(\ref{eq:Isigxxx}) at $T=300$~K. Arbitrary $y$ units.}
\label{fig:dfde}
\end{figure}
Fig.~\ref{fig:BiTe} demonstrates that potential problems at band crossings have negligible influence on the calculated transport coefficients for Bi$_2$Te$_3$ at 300~K. It should off-course be underlined that this is just one example and systems could exist where the present method should fail. However, band crossings only happen on symmetry lines (symmetry planes in hexagonal systems), so, while the problem exists, as long as the $\bk$-sampling is dense enough to keep the error localized at the crossing it will have little effect on global quantities like transport. Furthermore, at $T=300$~K, the $\partial f/\partial \varepsilon$ factor, Fig.~\ref{fig:dfde} is quite broad and has 5~\% of its maximum value at $\varepsilon-\mu=0.11$~eV. The transport coefficients are thus a sum over several Fermi surfaces and any problems will be smeared out.  

\begin{figure}[t]
\center
\includegraphics[width=.5\linewidth]{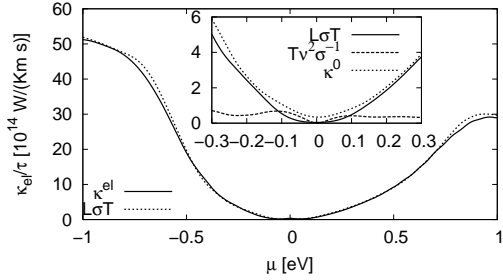} 
\caption{Electronic thermal conductivity as a function of chemical potential. Full line: calculated from Eqs.~(\ref{eq:k0}) and (\ref{eq:kel}). Dashed line: Calculated from Wiedemann-Franz law, Eq.~(\ref{eq:wf}) }
\label{fig:BiTe_2}
\end{figure}
As a further illustration and test of the method we have calculated the electronic thermal conductivity at zero electric current. This can be defined as
\beq
\kappa^{el}_{ij}=\kappa^0_{ij}-T\nu_{i\alpha}(\sigma^{-1})_{\beta\alpha}\nu_{\beta j}
\label{eq:kel}
\eeq
or aproximated through the electronic conductivity via Wiedemann-Franz law, which for degenerate charge carriers is given as
\begin{equation}
\kappa^{el}_{ij}= \frac{\pi^2}{3}\bigl( \frac{k_B}{e} \bigr)^2 \sigma_{ij} T 
\label{eq:wf}
\end{equation}
Figure~\ref{fig:BiTe}c shows the electronic thermal conductivity calculated with two different methods, Eqs.~(\ref{eq:kel}) and (\ref{eq:wf}). As expected the two lines are in very good agreement. The second term in Eq.~(\ref{eq:kel}), which is directly related to the power factor (Fig.~\ref{fig:BiTe}), is obviously insignificant far from the band-gap, where the Seebeck coefficient is small and the conductivity large. Close to the band gap it is a significant correction, as illustrated in the small insert in Fig.~\ref{fig:BiTe_2}.

\subsubsection{Test case: CoSb$_3$}
Calculations on CoSb$_3$ were carried out using the Engel-Vosko GGA.\cite{ev} The unit cell was $Im\bar{3}$ with $a=9.0385$~\AA. The Sb atom is placed at the $g$ Wyckoff position with (0,0.33537,0.15788). The plane-wave (PW) cut-off was defined by min($R_{MT}$)max($k_n$)=5.5 corresponding to approximately 588~PW. The Brillouin-zone (BZ) was sampled on a shifted tetrahedral mesh with 300~k points (17 in the IBZ) for the self consistent calculation. For the transport calculations a non-shifted mesh with 24000~k points (1030 in the IBZ) was used. The necessary derivatives were then calculated on a FFT grid five times as dense. 

\begin{figure}[t]
\includegraphics[width=\linewidth]{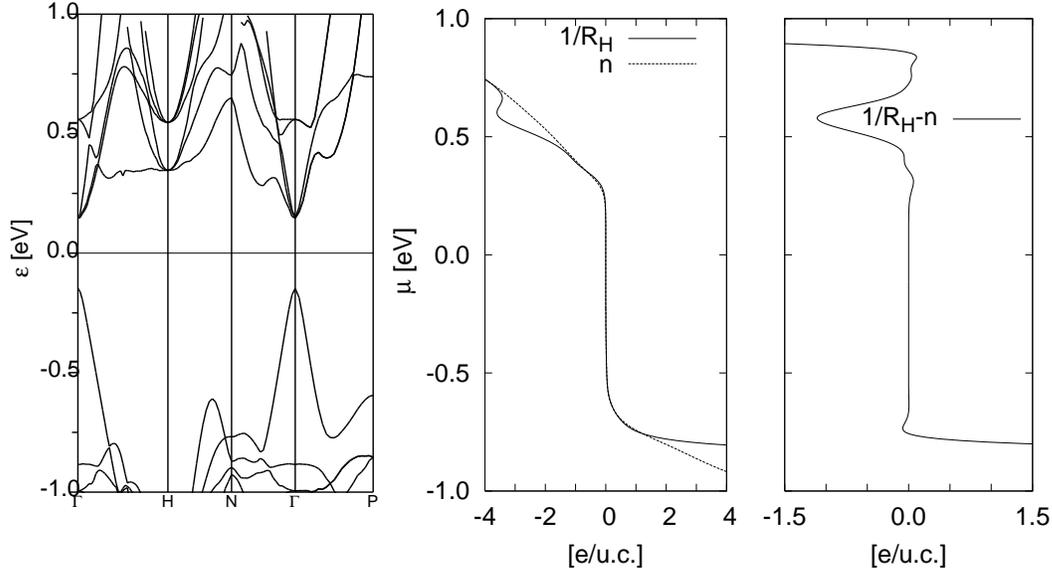} 
\caption{Band-structure of CoSb$_3$ together with the inverse Hall coefficient ($1/R_H$ and the calculated number of carriers $n$ (density of states) and the difference between $1/R_H$ and $n$.}
\label{fig:CoSb}
\end{figure}
The band structure of the skutterudite CoSb3 has been subject of some discussion and is sensitive to the lattice parameter and the exchange correlation function.\cite{davidskut1,sofoskut,kogaskut} It is generally agreed that it has parabolic bands close to the Fermi level, which we also find, Fig.~\ref{fig:CoSb}. In this region of parabolic bands the Hall coefficient should be inversely proportional to the number of carriers. The calculation of the Hall coefficient depends on the second derivatives of the bands, Eq.~(\ref{eq:sigxxx}), and therefore serves as a demanding test of the precision of the method and CoSb3 has therefore been chosen as a test example. Figure~\ref{fig:CoSb} illustrates how $1/R_H$ and the doping are almost equivalent in the region of parabolic bands, while they differ when the region of flat bands.

\section{Input parameters}
\begin{table}
\caption{Input variables. }
\begin{tabular}{l l}
\hline
\texttt{bandstyle}        &    Format of band structure to be input \\
%\texttt{iskip}            &    not used \\
\texttt{idebug}	          &    controls level of output\\
\texttt{eferm}            &    Fermilevel (in Ry)\\
\texttt{deltae}	          &    $d\varepsilon$ (in Ry)\\
\texttt{ecut}             &    cut-off energy around Fermilevel \\
%\texttt{smear-sigma}	  &    not used \\
\texttt{setgap}           &    logic switch for band gap manipulation \\
\texttt{gapsize}          &    new band-gap (in Ry) \\
\texttt{lpfac}            &    \# of times the interpolated mesh should be denser than the calculated \\
\texttt{efcut}            &    range of $\mu$ in which the integrations should be performed \\
\texttt{tmax}             &    max temperature at which the integrations should be performed \\
\texttt{deltat}           &    temperature step \\
%\texttt{Run type}         & CALC: perform band analysis. NOCALC: read transport distributions from file \\
\hline
\end{tabular}
\label{tab:input}
\end{table}
Table~\ref{tab:input} gives the input parameters used. \texttt{bandstyle} gives the format of the band-structure input. The present version of the code is interfaced to the band-structure code WIEN2k\cite{wien2k}, but can easily be interfaced to any other band-structure codes. $d\varepsilon$ defines how fine the mesh for the conductivity distribution should be, Eq.~(\ref{eq:transdist}). \texttt{ecut} defines the range of bands used around \texttt{eferm} in the integrals, Eqs.~(\ref{eq:Isigxx})-(\ref{eq:Isigxxx}). \texttt{setgap} and \texttt{gapsize} can be used to apply a scissors operator to force a certain band-gap. \texttt{lpfac} defines how much denser the interpolated mesh should be and thereby the $\bR$-cutoff in Eq.~\ref{eq:fourexp}. The programs outputs the conductivity tensors on a grid of $T$ and $\mu$, Eqs.~(\ref{eq:Isigxx})-(\ref{eq:Isigxxx}), defined by \texttt{efcut}, \texttt{tmax} and \texttt{deltat}. All output of the program is in SI-units

\section{Porting the code}
The present version of the code is interfaced to the band-structure code WIEN2k.\cite{wien2k} However, as the method uses only the crystal structure and the eigen-energies on a mesh as data the code is very easy to interface to other band-structure codes. The necessary crystal structure and band-structure information is contained in the \texttt{MODULE band-structure} and the user should therefore only supply a subroutine that sets up the module.

\begin{table}
\caption{The variables in \texttt{MODULE bandstructure}. }
\begin{tabular}{l l l}
\hline
\texttt{aac\_dir(3,3)}  &    REAL(8)              & Direction cosines for the conventional direct\\
\texttt{aac\_rec(3,3)}  &                         &  and reciprocal unit cell\\
\texttt{p2c\_dir(3,3)}  &    REAL(8)              & Conversion matrix for the primitive to conventional \\
\texttt{p2c\_rec(3,3)}  &                         & unit cell conversion\\
\texttt{nsym}             &    INTEGER            & number of symmetry operators\\
\texttt{symop(3,3,48)}	  &    REAL(8)              & symmetry operators with respect\\
 & & to the direct primitive lattice\\
\texttt{nband}		  &    INTEGER              & number of bands \\
\texttt{nkpt}		  &    INTEGER              & number of k-points in the IBZ \\
\texttt{xkpoint(:,:)}     &    REAL(8)              & $\bk$-points in basis of primitive reciprocal lattice vectors. \\
                          &                         &  Should be allocated as \texttt{xkpoint(3,nkpt)} \\
\texttt{bandenergy(:,:)}  &    REAL(8)              & eigen-energies in Ry.\\
                          &                         & Should be allocated as \texttt{bandenergy(nband,nkpt)} \\
\hline
\end{tabular}
\label{tab:moduleband}
\end{table}

\section{Conclusion}
We have implemented and tested a method for obtaining an analytical representation of the band-structure. We have applied it to the calculation of transport coefficients. The method has been compared with an earlier calculation\cite{sofote}, which in principle should be exact within Boltzmann theory, and we found a very convincing agreement.

It should be pointed out that the present method also has several advantages. First of all, when an analytical expression of the bands is found, they can be interpolated onto a finer $k$-mesh. Secondly, because only the energies are needed the code is easily portable to any band-structure code and furthermore does not require the storage of potentially large wavefunction files on disk. Finally, second derivatives necessary for the Hall coefficient are straightforwardly calculated which is not straightforward from the wavefunction itself. Even higher derivatives, necessary e.g. for the calculation of magneto-resistance, could easily be calculated with the present method, but it remains to be seen whether the accuracy is high enough.

%\bibliography{georg,solphys,te,vienna,clathrates}
%\bibliographystyle{elsart-num.bst}

\end{document}